%% file: 00_main.tex
\documentclass[letterpaper, 10 pt, conference]{ieeeconf}
\IEEEoverridecommandlockouts
\IEEEoverridecommandlockouts                              %

\usepackage{cancel}

\usepackage{amssymb}  %
\usepackage{mathtools}

\usepackage{import}
\usepackage{pgfplots}
\usepackage{color}
\usepackage{tabularx} %
\usepackage{multirow}
\usepackage{pgf}
\usepackage{blindtext}
\usepackage{todonotes}
\usepackage{cite}
\usepackage{siunitx}
\usepackage{balance}
\usepackage{epsfig}
\usepackage{comment}
\let\labelindent\relax
\usepackage{enumitem}
\usepackage{fancyhdr}

\def\new#1{{\textcolor{black}{{}#1}}}

\newcommand{\vect}[1]{\boldsymbol{#1}}

\renewcommand{\baselinestretch}{0.97}

\newlength{\bibitemsep}
\newlength{\bibparskip}
\let\oldthebibliography\thebibliography

\renewcommand\thebibliography[1]{%
  \oldthebibliography{#1}%
  \setlength{\parskip}{\bibitemsep}%
  \setlength{\itemsep}{\bibparskip}%
}
\AtBeginEnvironment{thebibliography}{\linespread{0.95}\selectfont}

\title{\LARGE \bf
Optimally Controlling the Timing of Energy Transfer in Elastic Joints:\\Experimental Validation of the Bi-Stiffness Actuation Concept
}

\author{Edmundo Pozo Fortuni\'{c}$^{1,*}$, Mehmet C. Yildirim$^{1,*}$, Dennis Ossadnik$^{1,*}$, Abdalla Swikir$^{1, 2}$, \\ Saeed Abdolshah$^{1}$, and Sami Haddadin$^{1}$ %
\thanks{$^{1}$The authors are with the Chair of Robotics and Systems Intelligence and the Munich Institute of Robotics and Machine Intelligence (MIRMI), Technical University of Munich, Germany. The authors acknowledge the financial support by the Bavarian State Ministry for Economic Affairs, Regional Development and Energy (StMWi) for the Lighthouse Initiative KI.FABRIK, (Phase 1: Infrastructure as well as the research and development program under, grant no. DIK0249) and European Union’s Horizon 2020 research and innovation programme as part of the project Darko under grant no. 101017274. Corresponding author: \tt\small{edmundo.pozo@tum.de}} 
\thanks{$^{2}$Abdalla Swikir is also with the Department of Electrical and Electronic Engineering, Omar Al-Mukhtar University (OMU), Albaida, Libya.} 
\thanks{$^{*}$The first three authors contributed equally to this work.}%
}

\makeatletter
 \let\old@ps@headings\ps@headings
 \let\old@ps@IEEEtitlepagestyle\ps@IEEEtitlepagestyle
 \def\confheader#1{%
 \def\ps@headings{%
 \old@ps@headings%
 \def\@oddhead{\strut\hfill#1\hfill\strut}%
 \def\@evenhead{\strut\hfill#1\hfill\strut}%
 }%
 \def\ps@IEEEtitlepagestyle{%
 \old@ps@IEEEtitlepagestyle%
 \def\@oddhead{\strut\hfill#1\hfill\strut}%
 \def\@evenhead{\strut\hfill#1\hfill\strut}%
 }%
 \ps@headings%
 }
 \makeatother
 
\begin{document}
\maketitle
\thispagestyle{fancy}
\lhead{\textit{This work has been submitted to the IEEE for possible publication. \\
Copyright may be transferred without notice, after which this version may no longer be accessible}}

\pagestyle{empty}

\input{01_introduction}

\input{02_concept}
\input{03_implementation}

\input{04_evaluation}
\input{05_conclusion}

\section*{Acknowledgement}
The authors would like to thank V. Rak\v{c}evi\'c, J. Ringwald, L. Chen and R. Franjga for the fruitful discussions.
	\balance
\bibliographystyle{ieeetr}

\bibliography{00_main.bib}

\end{document}

%% file: 01_introduction.tex
\begin{abstract}

\new{Elastic actuation taps into elastic elements' energy storage for dynamic motions beyond rigid actuation. While Series Elastic Actuators (SEA) and Variable Stiffness Actuators (VSA) are highly sophisticated, they do not fully provide control over energy transfer timing.} To overcome this problem on the basic system level, the Bi-Stiffness Actuation (BSA) concept was recently proposed. Theoretically, it allows for full link decoupling, while simultaneously being able to lock the spring in the drive train via a switch-and-hold mechanism. Thus, the user would be in full control of the potential energy storage and release timing. \new{In this work, we introduce an initial proof-of-concept of Bi-Stiffness-Actuation in the form of a 1-DoF physical prototype, which is implemented using a modular testbed. We present a hybrid system model, as well as the mechatronic implementation of the actuator. We corroborate the feasibility of the concept by conducting a series of hardware experiments using an open-loop control signal obtained by trajectory optimization.} Here, we compare the performance of the prototype with a comparable SEA implementation. We show that BSA outperforms SEA 1) in terms of maximum velocity at low final times and 2) in terms of the movement strategy itself: The clutch mechanism allows the BSA to generate consistent launch sequences while the SEA has to rely on lengthy and possibly dangerous oscillatory swing-up motions. Furthermore, we demonstrate that providing full control authority over the energy transfer timing and link decoupling allows the user to synchronously release both elastic joint and gravitational energy. This facilitates the optimal exploitation of elastic and gravitational potentials in a synergistic manner.
\end{abstract}
\section{Introduction}
Intrinsically elastic joints have been introduced to the robotics community several decades ago. The capability to store potential energy in springs allows not only for more efficient robot motion but also the generation of explosive movements \cite{braun2012exploiting} achieving (super-)human performance \cite{haddadin2012intrinsically}. Optimally controlling the energy flow can reduce the energy consumption of the system \cite{nieto2019minimizing} or lead to an increase in output power \cite{haddadin2011optimal}. Another important property of intrinsically elastic joints is the increased safety level in scenarios where a collision between human and robot is possible \cite{park2009safe}. An elastic arm operates more safely than a similar rigid joint robot at the same velocity \cite{mansfeld2021speed}. 

Among the different types of elastic joint architectures, \textit{Series Elastic Actuation} (SEA) is one of the most widely used, as seen in \cite{mcy_sea_2021}. Here, an elastic element is connected between the actuator and the link. It has been shown that by exploiting the natural modes of the system via a bang-bang-like resonant excitation signal \cite{haddadin2009kick}, much higher velocities can be achieved when compared to the rigid joint case. This can be utilized to create highly dynamic kicking, throwing, or jumping motions \cite{icub_jumping, haddadin2009kick, Haddadin2012}. However, it is not possible to independently control the timing of energy storage and release, which limits the usability of such systems \cite{Plooij}.

For better control of the energy flow in the actuator, various efforts have been made to include clutches in the drive train. A wide variety of combinations of elastic and coupling elements exists, which can be summarized under the umbrella term \textit{Clutched Elastic Actuation} (CEA) \cite{Plooij}. In \cite{deboon2019differentially}, a differential clutch was added to the SEA for a robot-assisted rehabilitation scenario and despite the generated complexity, it provided an increased level of controllability. Clutched SEAs are also used in the development of prostheses and reduced energy consumption \cite{convens2019modeling,rouse2014clutchable}. 
\begin{figure}[t]
    \begin{minipage}{\linewidth}
    \centering
          	\def\svgwidth{0.7\linewidth}
	\input{./figures/setup_new.pdf_tex}
\end{minipage}
\begin{minipage}{\linewidth}
\mbox{\;\;\;\;\;(c)}
\end{minipage}
\begin{minipage}{\linewidth}
\centering
\vspace{-0.2cm}
  	\def\svgwidth{0.9\linewidth}
	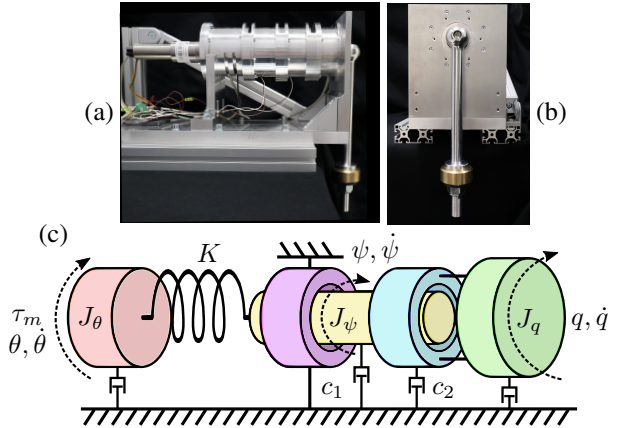 
\end{minipage}
    \vspace{-0.2cm}
    \caption{Hardware setup (a)-(b) and ideal  system model (c). The motor (red) is connected to the spring inertia (yellow), which can be either braked (violet clutch) or connected to the link (green) by the cyan clutch.}    
    \label{fig:ideal_model}
    \vspace{-0.7cm}
\end{figure} 

Another way to mitigate the dependency of energy storage and release timing is the use of \textit{Variable Stiffness Actuation} (VSA). In \cite{Haddadin2012}, using optimal control, it was possible to maximize the energy transfer from the proximal to the distal links by an approximate decoupling of the link from the motor. As a result, clear launch sequences instead of resonant swing-up motions emerged for short end-times. The trajectories closely resembled the proximo-distal sequence observed in biomechanics, where each link in the chain contributes at exactly the right time in order to maximize the energy transfer -- a phenomenon sometimes called \textit{inertia timing} \cite{grezios2006muscle}. This strategy requires two distinct stiffness modes to generate the proximo-distal sequence:  \new{ A high stiffness setting (corresponding to the maximum stiffness of the VSA) and a very low stiffness setting. However, a clear proximo-distal sequence only emerges when carefully selecting the end-time in the optimal control formulation: As shown in our previous work \cite{ossadnik2022BSA}, using higher-end times in the optimal control formulation results in resonant excitation as the prevalent strategy, similar to SEA.}

 Previously, in \cite{ossadnik2022BSA}, we introduced a novel CEA concept denoted \textit{Bi-Stiffness Actuation} (BSA). The concept consists of a motor attached to a spring and a switch-and-hold mechanism. \new{This mechanism offers two distinct modes: one can brake the spring element and another couples it to the link side.  This corresponds to a pre-defined stiffness and a zero stiffness mode. This setup shares similarities with the ideal VSA with one notable distinction: The zero stiffness setting replaces the low-stiffness setting in the ideal VSA. Thus, BSA is able to achieve full link decoupling.} Additionally, the spring can be restrained while the link is disengaged from the drive train, allowing it to store potential energy independently of the link's motion. So far, the concept was only studied in simulation and no physical proof-of-concept has been provided.
\subsection{Contribution}
Therefore, in this work, we introduce a 1-DoF physical prototype of the actuator concept suggested in \cite{ossadnik2022BSA}. The abstract switch-and-hold mechanism is interpreted as two individual clutches (see Fig.~\ref{fig:ideal_model}). We also present the hybrid system model as well as the mechatronic implementation of the actuator. For evaluation, we conducted a series of hardware experiments that compare the performance of the BSA prototype with a comparable SEA. Using optimal control, we demonstrate that
\begin{itemize}[leftmargin=*]
    \item  Similar to a trebuchet, the actuator can optimally brake and release the spring element for precisely timing the energy transfer.
    \item Using the above strategy, a higher link velocity can be reached in less time and in a much more controlled fashion in comparison to SEA, which has to rely on lengthy and possibly dangerous oscillatory swing-up motions to achieve similar speeds.
    \item Decoupling can be used for optimally exploiting the synchronous release of both elastic and gravitational energy; again leading to higher speeds in comparison to the SEA in a limited time frame. 
\end{itemize}
We not only validate the simulation results (maximization of the link velocity at different end-times) given in~\cite{ossadnik2022BSA} by conducting experiments on the actual prototype, but we also provide an in-depth view of the mechanic and electronic implementation of such a system. Moreover, we show the differences between ideal and implementation parameters and further validate the BSA concept in an additional scenario (maximization of the link velocity with varying initial link positions).
In the remainder of this paper, details of the BSA concept and its modelling are given in Sec.~\ref{sec:concept_modelling}. The implementation of the concept in a prototype is presented in Sec.~\ref{sec:implementation}. Then, the prototype is evaluated in Sec.~\ref{sec:evaluation}. Finally, the manuscript is concluded in Sec.~\ref{sec:conclusion} by discussing the results and addressing possible future research directions.

%% file: figures/setup_new.pdf_tex
\begingroup%
  \makeatletter%
  \providecommand\color[2][]{%
    \errmessage{(Inkscape) Color is used for the text in Inkscape, but the package 'color.sty' is not loaded}%
    \renewcommand\color[2][]{}%
  }%
  \providecommand\transparent[1]{%
    \errmessage{(Inkscape) Transparency is used (non-zero) for the text in Inkscape, but the package 'transparent.sty' is not loaded}%
    \renewcommand\transparent[1]{}%
  }%
  \providecommand\rotatebox[2]{#2}%
  \newcommand*\fsize{\dimexpr\f@size pt\relax}%
  \newcommand*\lineheight[1]{\fontsize{\fsize}{#1\fsize}\selectfont}%
  \ifx\svgwidth\undefined%
    \setlength{\unitlength}{7739.44289801bp}%
    \ifx\svgscale\undefined%
      \relax%
    \else%
      \setlength{\unitlength}{\unitlength * \real{\svgscale}}%
    \fi%
  \else%
    \setlength{\unitlength}{\svgwidth}%
  \fi%
  \global\let\svgwidth\undefined%
  \global\let\svgscale\undefined%
  \makeatother%
  \begin{picture}(1,0.48737294)%
    \lineheight{1}%
    \setlength\tabcolsep{0pt}%
    \put(0,0){\includegraphics[width=\unitlength,page=1]{./figures/setup_new.pdf}}%
    \put(-0.033,0.22895335){\color[rgb]{0,0,0}\makebox(0,0)[lt]{\lineheight{1.25}\smash{\begin{tabular}[t]{l}(a)\end{tabular}}}}%
    \put(0.96,0.22895335){\color[rgb]{0,0,0}\makebox(0,0)[lt]{\lineheight{1.25}\smash{\begin{tabular}[t]{l}(b)\end{tabular}}}}%
  \end{picture}%
\endgroup%

%% file: figures/BSA_annotated.pdf_tex
\begingroup%
  \makeatletter%
  \providecommand\color[2][]{%
    \errmessage{(Inkscape) Color is used for the text in Inkscape, but the package 'color.sty' is not loaded}%
    \renewcommand\color[2][]{}%
  }%
  \providecommand\transparent[1]{%
    \errmessage{(Inkscape) Transparency is used (non-zero) for the text in Inkscape, but the package 'transparent.sty' is not loaded}%
    \renewcommand\transparent[1]{}%
  }%
  \providecommand\rotatebox[2]{#2}%
  \newcommand*\fsize{\dimexpr\f@size pt\relax}%
  \newcommand*\lineheight[1]{\fontsize{\fsize}{#1\fsize}\selectfont}%
  \ifx\svgwidth\undefined%
    \setlength{\unitlength}{209.68140369bp}%
    \ifx\svgscale\undefined%
      \relax%
    \else%
      \setlength{\unitlength}{\unitlength * \real{\svgscale}}%
    \fi%
  \else%
    \setlength{\unitlength}{\svgwidth}%
  \fi%
  \global\let\svgwidth\undefined%
  \global\let\svgscale\undefined%
  \makeatother%
  \begin{picture}(1,0.34736841)%
    \lineheight{1}%
    \setlength\tabcolsep{0pt}%
    \put(0,0){\includegraphics[width=\unitlength,page=1]{./figures/BSA_annotated.pdf}}%
    \put(-0.03927392,0.19542491){\color[rgb]{0,0,0}\makebox(0,0)[lt]{\lineheight{1.25}\smash{\begin{tabular}[t]{l}$\tau_m$\end{tabular}}}}%
    \put(-0.04438375,0.13998373){\color[rgb]{0,0,0}\makebox(0,0)[lt]{\lineheight{1.25}\smash{\begin{tabular}[t]{l}$\theta, \dot{\theta}$\end{tabular}}}}%
    \put(0.54188275,0.30220507){\color[rgb]{0,0,0}\makebox(0,0)[lt]{\lineheight{1.25}\smash{\begin{tabular}[t]{l}$\psi, \dot{\psi}$\end{tabular}}}}%
    \put(0.91803291,0.18790917){\color[rgb]{0,0,0}\makebox(0,0)[lt]{\lineheight{1.25}\smash{\begin{tabular}[t]{l}$q, \dot{q}$\end{tabular}}}}%
    \put(0.27767451,0.29652904){\color[rgb]{0,0,0}\makebox(0,0)[lt]{\lineheight{1.25}\smash{\begin{tabular}[t]{l}$K$\end{tabular}}}}%
    \put(0,0){\includegraphics[width=\unitlength,page=2]{./figures/BSA_annotated.pdf}}%
    \put(0.82300829,0.18422056){\color[rgb]{0,0,0}\makebox(0,0)[lt]{\lineheight{1.25}\smash{\begin{tabular}[t]{l}$J_q$\end{tabular}}}}%
    \put(0.50255293,0.18573899){\color[rgb]{0,0,0}\makebox(0,0)[lt]{\lineheight{1.25}\smash{\begin{tabular}[t]{l}$J_{\psi}$\end{tabular}}}}%
    \put(0.07329649,0.18765266){\color[rgb]{0,0,0}\makebox(0,0)[lt]{\lineheight{1.25}\smash{\begin{tabular}[t]{l}$J_{\theta}$\end{tabular}}}}%
    \put(0.49,0.07021346){\color[rgb]{0,0,0}\makebox(0,0)[lt]{\lineheight{1.25}\smash{\begin{tabular}[t]{l}$c_1$\end{tabular}}}}%
    \put(0.68,0.07021346){\color[rgb]{0,0,0}\makebox(0,0)[lt]{\lineheight{1.25}\smash{\begin{tabular}[t]{l}$c_2$\end{tabular}}}}%
  \end{picture}%
\endgroup%

%% file: 02_concept.tex
\section{Modelling} 
\label{sec:concept_modelling}

In this section, we first present the architecture of the actuator in Sec.~\ref{sec:concept}. Then, we present an idealized model of the system in Sec.~\ref{sec:modelling}.
\subsection{High-level architecture} \label{sec:concept}
We introduce $\theta$, $\psi$ and $q$, which correspond to the motor, spring and link position, respectively. The associated mass moments of inertia $J$ are marked with their corresponding subscripts. The different modes of the actuator are implemented by engaging and disengaging two mechanical clutches, $c_1$ and $c_2$, which are used to brake the inertia of the spring (that is, connecting it to a fixed frame) and coupling it to the link, respectively. The main advantage of the actuator is the direct control over the energy transfer timing, since the spring can be loaded while the link is moving due to inertial coupling or gravity. 

\subsection{Modelling} \label{sec:modelling}
In the following, we derive the hybrid system model of our actuator. Hybrid systems are characterized by the interplay between continuous and discrete dynamics. We will derive expressions for these subcomponents one by one before we form the full hybrid system model at the end of this section. %

\textbf{Equation of motion.} We introduce the vector $\vect{\xi} = [\psi, q]^\mathsf{T}$ and state the actuator's equation of motion
\begin{gather}
    \underbrace{\begin{bmatrix}
        J_{\psi} & 0 \\ 0 & J_q
    \end{bmatrix}}_{\eqqcolon \vect{B}} \Ddot{\vect{\xi}} +  \underbrace{\begin{bmatrix}
        0 \\
         m g l \sin (q)
    \end{bmatrix}}_{\eqqcolon \vect{g}(\vect{\xi})} 
    +  \vect{\tau}_{f} = \underbrace{\begin{bmatrix}
        \tau_s \\
        0
    \end{bmatrix}}_{\eqqcolon \vect{\tau}_s}  + \vect{C}_p^{\mathsf{T}} \vect{\lambda} \label{eq:dynamics}\\
    J_{\theta} \Ddot{\theta} + \tau_s = \tau_m \label{eq:motor_dyn}  \\
    \vect{C}_p \dot{\vect{\xi}} = \vect{0}. \label{eq:constr}
\end{gather}
with the constant mass matrix $\vect{B}$, the gravitational vector $\vect{g}(\vect{\xi})$, the motor torque $\tau_m$ and the friction torque $\vect{\tau}_f$. The matrix $\vect{C}_p$ defines the switching through the clutch mechanism and $\vect{\lambda}$ is the constraint torque. We assume that the motor coupling acts solely through the spring (cf. \cite{spong1989modeling}). The spring torque is given by $\tau_s = K \phi$, where  $\phi \coloneqq \theta -\psi$ is the deflection\footnote{The real system exhibits some hysteresis behaviour. However, to simplify the optimal control formulation, we still assume a linear spring model. Hysteresis can be modelled by increasing the spring torque by a loss term $h$, which yields $\tau_s = K (\phi-h)$. The dynamics of the nonlinear loss term can be described by the Bouc-Wen model~\cite{song2006generalizedboucwen} as $\dot{h} = \alpha \dot{\phi} - \beta |\phi| h - \gamma \dot{\phi} |h|$, where $\alpha$, $\beta$, and $\gamma$ are the non-dimensional shaping factors for the hysteresis model.}.

\textbf{Clutch torque.} The role of the clutch in the actuator is to prevent relative motion between frames. If the spring is braked, $\dot \psi = 0$ holds. Similarly, if the link is engaged, $\dot \psi = \dot q$. These conditions are encoded in the matrices $\vect{C}_p$ and summarized in Tab.~\ref{tab:modes}. Both clutches can be active or inactive at the same time. Thus, there are two additional modes compared to the system introduced in \cite{ossadnik2022BSA}. To solve for the constraint torque $\vect{\lambda}$, we differentiate \eqref{eq:constr}, which yields $\vect{C}_p \ddot{\vect{\xi}} = \vect{0}$, since $\dot{\vect{C}}_p = 0$. Substituting \eqref{eq:dynamics} in \eqref{eq:constr} leads to
\begin{equation} \label{eq:constr_torque}
		 \vect{\lambda} = (\vect{C}_p \vect{B}^{-1} \vect{C}_p^{\mathsf{T}})^{-1}\vect{C}_p \vect{B}^{-1} (\vect{g}(\vect{\xi})+ 
        \vect{\tau}_{f}
    - \vect{\tau}_s ).
\end{equation} 
For each mode $p$, we now have obtained an explicit expression for the clutch torque that has to act to satisfy the constraints. Next, we show how to deal with switching between modes.

\textbf{Impact analysis.} When the contact situation changes, i.e., the clutch is engaged or disengaged, the new constraint is enforced by an impulse. We assume that this impulse leads to an instantaneous change in the velocity of the system, leading to the following update law:
\begin{align}\label{eq:impact_law}
	\dot{\vect{\xi}}^{+} &= \dot{\vect{\xi}}^{-} +  \vect{B}^{-1} \vect{C}_p^{\mathsf{T}} \vect{\Lambda}_p, \\ 
	\vect{\Lambda}_p &= -(\vect{C}_p \vect{B}^{-1} \vect{C}_p^{\mathsf{T}})^{-1} \vect{C}_p \dot{\vect{\xi}}^{-},    
\end{align}

\noindent where $\vect{\Lambda}_p$ is the contact impulse. The superscripts $^-$ and $^+$ denote the vector before and after the impact, respectively. A full derivation can be found in \cite{ossadnik2022BSA, ossadnik2021nonlinear}.

\textbf{Friction modelling.} The friction torque due to the bearings is modelled by a Coulomb and viscous friction term
\begin{equation}
    \vect{\tau}_f = \begin{bmatrix}
        \tau_{C, \psi} \mathsf{sign}(\psi) - d_{\psi} \dot{\psi} \\
        \tau_{C, q} \mathsf{sign}(q) - d_q \dot{q} 
    \end{bmatrix}.
\end{equation}

\noindent Here, $\tau_{C, \psi}$ and $\tau_{C, q}$ are the constant Coulomb friction terms for the spring- and link-side and $d_{\psi}$ and $d_{q}$ are viscous damping terms, respectively. %
Before we put everything together in the hybrid system model, we present a method to simplify the motor dynamics. 

\textbf{Singular perturbation.} Given a control law \mbox{$\tau_m = k_p (\dot \theta_d - \dot \theta)$}, 
the motor dynamics \eqref{eq:motor_dyn} can be brought in singular perturbation form (cf. \cite{Haddadin2012})
\begin{equation}
   \epsilon (J_m \Ddot{\theta} + \tau) =  \dot \theta_d - \dot \theta,
\end{equation}
\noindent where $\epsilon = 1/ k_p$. Taking the limit $\epsilon \rightarrow 0$ leads to $ \dot \theta = \dot \theta_d$. We can now define $u \coloneqq \dot \theta_d$ as the new control input. Now, we have all the elements that are necessary to create the hybrid system model.

\textbf{Hybrid system model.} %
Since the mode transition is characterized by an instantaneous change of velocity, our system falls into the category of switched-impulsive systems (cf. \cite{liberzon2003switching}). We define the continuous state $\vect{x} \coloneqq [\theta, \vect{\xi}, \Dot{\vect{\xi}} 
]^\mathsf{T} \in \mathbb{R}^5$ and the index set $\mathcal{P} \coloneqq \{0, 1, 2, 3\}$. Continuous dynamics are defined by a family of vector fields
\begin{equation}
	\dot{\vect{x}} = \vect{f}_p(\vect{x}, u) \coloneqq \begin{bmatrix}
	u \\
	\dot{\vect{\xi}} \\
	 \vect{B}^{-1} (\vect{C}_{p}^{\mathsf{T}} \vect{\lambda}_{p} +\vect{\tau}_s- \vect{g}(\vect{\xi}) - \vect{\tau}_f)
	\end{bmatrix}, \label{eq:dyn}
\end{equation}
\noindent where $p \in \mathcal{P}$. After a mode transition, the state of the system has to be reinitialized by evaluating a reset map, which is defined by
\begin{equation}
    \vect{x}^{+} = \vect{g}_p(\vect{x}^{-}) \coloneqq \begin{bmatrix}
    \theta^{-} \\
    \vect{\xi}^{-} \\
    \dot{\vect{x}}^{-} + \vect{B}^{-1} \vect{C}_{p}^{\mathsf{T}} \vect{\Lambda}_{p} \\
    \end{bmatrix}. \label{eq:jump}
\end{equation}
\noindent Given \eqref{eq:dyn} and \eqref{eq:jump}, we define an external switching signal \mbox{$\sigma: \mathbb{R}^{+} \rightarrow \mathcal{P}$} in such a way that we create a switched-impulsive system
\begin{align}
	\dot{\vect{x}} = \vect{f}_{\sigma}(\vect{x}, \vect{u}),~
	\vect{x}^{+} = \vect{g}_{\sigma}(\vect{x}^{-}).\label{eq:dyn1}
\end{align}
Please note that the values of all mechanical parameters from all previous equations are listed in Table \ref{tab:parameters}, Sec.~\ref{sec:evaluation}.
\input{tables/cases.tex}

%% file: tables/cases.tex
\begin{table}[t]
\vspace{0.2cm}
\caption{Actuator modes. DEC (Decoupled), SEA, STG (Storage), BRK (Braked Storage).  If $c_i = 1$ the clutch is engaged, else if  $c_i = 0$ it is disengaged.} \label{tab:modes}
\vspace{-0.6cm}
\begin{minipage}{\linewidth}
\begin{minipage}{0.65\linewidth}
\centering

    \begin{tabular}[c]{|c|c|c|c|c|} 
	\hline	\textbf{Mode} & $p$& $c_1$ & $c_2$ &  $\vect{C}_p$\\
	\hline \rule{0pt}{4ex}    DEC &0 &0& 0& $- $        ~\\[0.25cm]
	\hline	\rule{0pt}{4ex}    SEA &1&0& 1& $\begin{bmatrix}
	1& -1
	\end{bmatrix}$ ~\\[0.25cm]
 	\hline \rule{0pt}{4ex}    STG &2 &1& 0& $ \begin{bmatrix}
	1 & 0
	\end{bmatrix}$        ~\\[0.25cm]
 	\hline \rule{0pt}{4ex}    BRK &3 &1& 1& $ \begin{bmatrix}
	1 & 0 \\
 	1 & -1

	\end{bmatrix}$        ~\\[0.25cm]
 \hline
\end{tabular}
\end{minipage} \begin{minipage}{0.34\linewidth}
 \def\svgwidth{1.\linewidth}
 \vspace*{0.05cm}
	\hspace*{-0.3cm} 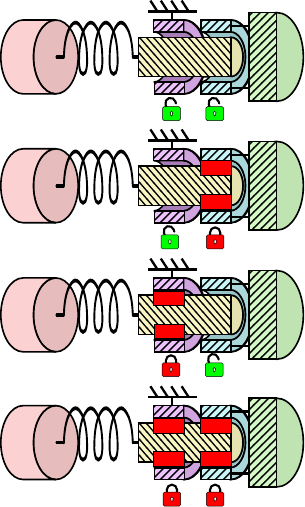 
\end{minipage}

\end{minipage}
\vspace{-0.7cm} 
\end{table}

%% file: figures/SEA_DEC.pdf_tex
\begingroup%
  \makeatletter%
  \providecommand\color[2][]{%
    \errmessage{(Inkscape) Color is used for the text in Inkscape, but the package 'color.sty' is not loaded}%
    \renewcommand\color[2][]{}%
  }%
  \providecommand\transparent[1]{%
    \errmessage{(Inkscape) Transparency is used (non-zero) for the text in Inkscape, but the package 'transparent.sty' is not loaded}%
    \renewcommand\transparent[1]{}%
  }%
  \providecommand\rotatebox[2]{#2}%
  \newcommand*\fsize{\dimexpr\f@size pt\relax}%
  \newcommand*\lineheight[1]{\fontsize{\fsize}{#1\fsize}\selectfont}%
  \ifx\svgwidth\undefined%
    \setlength{\unitlength}{146.00951187bp}%
    \ifx\svgscale\undefined%
      \relax%
    \else%
      \setlength{\unitlength}{\unitlength * \real{\svgscale}}%
    \fi%
  \else%
    \setlength{\unitlength}{\svgwidth}%
  \fi%
  \global\let\svgwidth\undefined%
  \global\let\svgscale\undefined%
  \makeatother%
  \begin{picture}(1,1.66439464)%
    \lineheight{1}%
    \setlength\tabcolsep{0pt}%
    \put(0,0){\includegraphics[width=\unitlength,page=1]{./figures/SEA_DEC.pdf}}%
  \end{picture}%
\endgroup%

%% file: 03_implementation.tex
\section{Implementation}
\label{sec:implementation}
 One of the main contributions of this paper is the physical implementation of the BSA concept. This allows us to not only demonstrate its feasibility but also conduct experiments to show the advantages of the actuator over conventional systems such as SEAs. For this purpose, we developed a configurable modular testbed. The modules can be combined to produce the BSA configuration explained in the previous section. Additionally, each module, except the motor, can potentially be connected to any other module sequentially and in any order, including additional instances of the same type. As a result, this allows not only for the implementation of, for example, a BSA or SEA configuration, but also a broad number of other mechanisms. The modular setup also allows us to operate and test the effects of each element independently, such as link decoupling, the performance of the elastic element, stiffness brake and motor control. This, in turn, leads to a challenging list of design and implementation requirements. In this section, we present the essential steps of the system development, followed up by the necessary identification processes. We provide a comprehensive description of the specific hardware components that were used, enabling anyone to replicate our design.
\begin{figure*}[t]
\vspace{0.25cm}
    \centering
    \includegraphics[width =\textwidth]{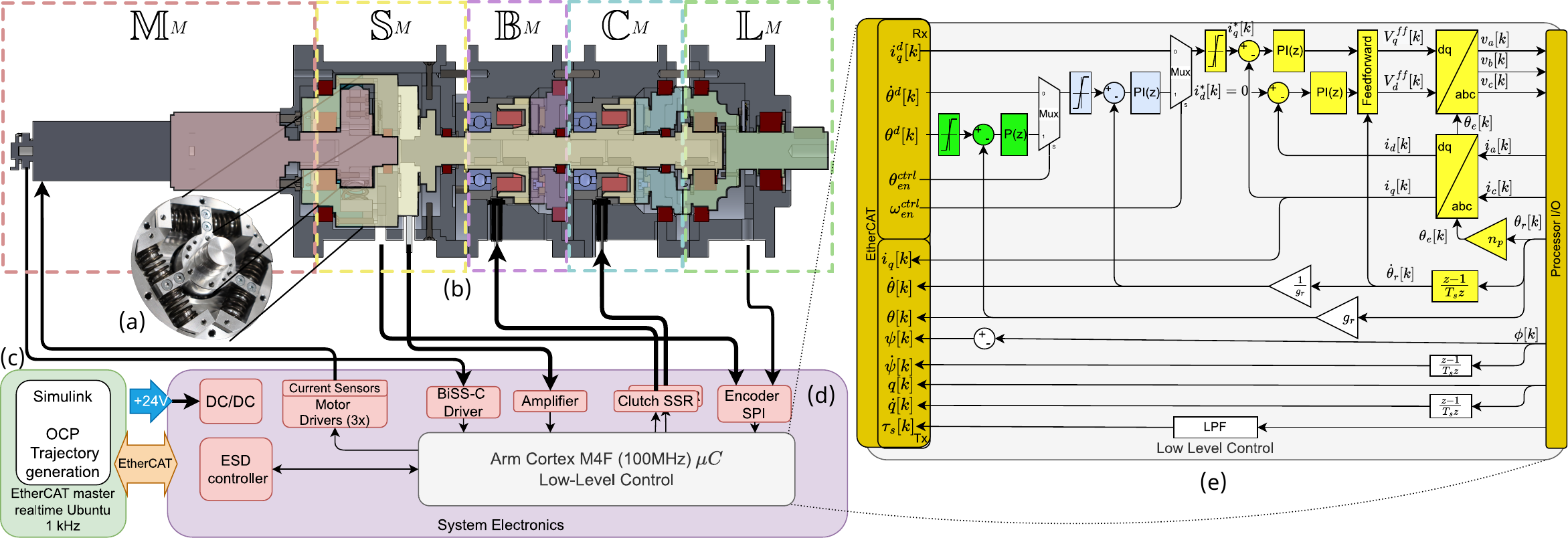}
    \vspace{-0.6cm}
    \caption{Detailed design of the implemented configurable modular testbed in BSA configuration ($\mathbb{M}^{M}$-$\mathbb{S}^{M}$-$\mathbb{B}^{M}$-$\mathbb{C}^{M}$-$\mathbb{L}^{M}$). (a) Designed spring element. (b) Section-view of CAD assembly. (c) PC-based main controller. (d) Custom-made electronics. (e) Low-level Cascaded Controller.}
    \label{fig:prototype_section_view}
    \vspace{-0.6cm}
\end{figure*}

\subsection{Architecture and Mechanical Design}
\label{sec:architecture}

\begin{figure}[t]
  \centering
  	\def\svgwidth{\linewidth}
   \tiny{
	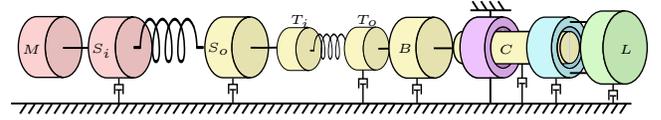 }
        \vspace{-0.4cm}
    \caption{Implementation model of the BSA concept. New elements are added to reflect the real components of the modular testbed. Thus, it can be converted to its equivalent model, as in Fig.~\ref{fig:ideal_model}(c).}
    \label{fig:implementation_model}
    \vspace{-0.5cm}
\end{figure}

In the mechanical design process, we focused on two main paradigms: 1) Maximum modularity, 2) Use of off-the-shelf components to limit the potential error caused by the custom hardware.
\new{The system consists of five types of modules:}  motor ($\mathbb{M}^{M}$), spring ($\mathbb{S}^{M}$), brake ($\mathbb{B}^{M}$), clutch ($\mathbb{C}^{M}$), and link ($\mathbb{L}^{M}$). %
All modules are designed with compatible $a^f, b^f, c^f$ inputs and $a^m,b^m,c^m$ outputs, where the superscripts $^f$ and $^m$ denote a female and male connection, respectively. For example, with a combination of $\mathbb{M}^{M}$-$\mathbb{C}^{M}$-$\mathbb{S}^{M}$ a similar system as suggested in~\cite{compact_clutched2013} can be \new{obtained. In the following, we will provide a thorough breakdown of each module and outline the design choices made for each.}
\begin{itemize}[leftmargin=*]
    \item $\mathbb{M}^{M}$ consist of a 3274G024BP4-6356 BLDC motor (Faulhaber GmbH, Germany) with embedded AEMT-12/16L magnetic multi-turn absolute encoder and 42GPT 1:108 planetary gearbox and a manufactured adapter part to connect to the rigid frame of the other modules. \new{This combination was selected due to its high power-to-weight ratio and wide operational range.} The module has an $a^{m}$-type output.
    \item $\mathbb{S}^{M}$ consists of an inner body with two inputs ($a^{f}$ and $b^{f}$) and one output shaft ($a^{m}$) that contains a rotational spring based on Tsagarakis et al.~\cite{compact_sea2009} (See Fig.~2(a).), an embedded single turn AksIM-2 MB039SPL19BENT00 (RLS d.o.o, Slovenia) absolute magnetic encoder and a TS70a-5Nm (ME-Meßsystem GmbH, Germany) torque sensor for spring characterization and monitoring. The advantage of this design lies in the use of off-the-shelf linear springs \new{allowing the implementation of multiple stiffnesses. We aimed to select our stiffness according to the following two criteria: 1) Motion range and 2) energy efficiency. According to the guidelines in \cite{Verstraten2016} and given the mechanical constraints of the architecture, $K = 15$ Nm/rad is selected as a good trade-off for a wide motion range at a reasonable cost in terms of energy efficiency. Using this stiffness, we can also exploit the full deflection range within the given torque sensor limit.}
    
    \item $\mathbb{B}^{M}$ and $\mathbb{C}^{M}$ are developed using a normally disengaged tooth clutch 546.12.3.4 (Mönninghoff, Germany), each. On $\mathbb{B}^{M}$, the clutch has a rigid connection between its $a^{f}$ input and $a^{m}$ output shafts and toggles a connection to the outer frame of the module, braking and releasing the system, while the clutch in $\mathbb{C}^{M}$ toggles the connection of its $b^{m}$ output to its rigid connection between the $a^{f}$ input and $a^{m}$ output shafts. Contrary, for example, to friction disc clutches, a tooth clutch allows for a rapid change of the contact mode and thus realizes the instantaneous switching that we presuppose in the hybrid system formulation.
    \item  $\mathbb{L}^{M}$ consists of a series of cross-roller bearings and an embedded absolute magnetic encoder, as in $\mathbb{S}^{M}$. Its $b^{f}$ input is connected to its $c^{m}$ output and its $a^{f}$ input is connected to a free-running bearing. The $\mathbb{L}^{M}$ can also be assembled to have its $a^{f}$ input rigidly connected to its output too, providing some additional flexibility in possible future developments.
\end{itemize}
\vspace{-0.02cm}
To construct a BSA configuration, a permutation of $\mathbb{M}^{M}$-$\mathbb{S}^{M}$-$\mathbb{B}^{M}$-$\mathbb{C}^{M}$-$\mathbb{L}^{M}$ is assembled as depicted in Fig.~\ref{fig:ideal_model}.(a)-(b). The section view of the CAD assembly can be seen in Fig.~\ref{fig:prototype_section_view}.(b). The rigidly connected parts are highlighted with the same colour code seen in Fig.~\ref{fig:ideal_model} to maintain consistency. By doing so, the implementation model of the BSA, which reflects the real components present in the assembled actuator, is achieved, as depicted in Fig.~\ref{fig:implementation_model}. The relationship between the parameters of the modular-based implementation model and the ideal model parameters can be seen in Table~\ref{tab:parameters}. Given that the definitions of the inertial bodies and the external disturbances are defined according to each module, the resultant values for the different permutations can be calculated in a simple manner.

\subsection{Electronics and Control}
\label{sec:electronics_control}

The software, electronics, and control architecture are depicted in Fig.~\ref{fig:prototype_section_view}(c)-(e). \\
\textbf{Main Controller.} For quick employment of our optimal controllers (in~\ref{sec:optimal_control}), we use an EtherCAT framework (cf. Fig.~\ref{fig:prototype_section_view}(c)). A control PC (x64 running an Ubuntu 20.04) hosts the high-level control routine using a Matlab / Simulink environment (MathWorks, MA, USA). It also hosts an EtherCAT master controller using Etherlab (Ingenieurgemeinschaft IgH, Germany) at 1 kHz, including the submodule for interfacing the EtherCAT slave devices (ESD). \\
\textbf{Embedded Controller Hardware.} To reduce the electronic footprint, we developed a custom low-level slave device subsystem, composed of a custom-made embedded brushless DC motor controller board, as shown in Fig.~\ref{fig:prototype_section_view}(d). It is based on an ARM Cortex-M4F microcontroller ($\mu$C) with a single-precision floating-point unit (FPU) running at 100 MHz. Two ACS71240KEXBLT-010B3 giant magneto-resistant (GMR)-based current sensors are located in line to phases A and C of the motor and connected to two independent 12bit analogue-to-digital converters (ADC) to read currents $i_a$ and $i_c$, respectively. A strain gauge Wheatstone full-bridge torque sensor differential amplifier circuit is connected to an additional channel of an ADC to read $\tau_s$ from $\mathbb{S}^{M}$. One Serial-Peripheral-Interface (SPI) bus is used to communicate with the magnetic encoders and read $\phi$ and $q$ from $\mathbb{S}^{M}$ and $\mathbb{L}^{M}$, respectively, and an additional SPI bus with a BiSS-C transceiver is used to receive data from the BLDC rotor position, $\theta_{r}$. Both the ADCs and SPI modules are synchronized with the Pulse Width Modulation (PWM) module used for driving the motor via three half-Bridge inverters. A 12.5kHz sampling and PWM frequency is selected as the best alternative considering: PWM resolution, sensor sampling frequency and the execution of the embedded controller firmware. Additionally, two digital channels are provided to turn on/off up to two clutch/brake actuators using CPC1002N miniature DC solid state relays (SSR). The system also comprises a real-time EtherCAT-based slave communication sub-module running at 1kHz using a LAN9253 slave controller.\\
\begin{figure}[t]
    \vspace{0.2cm}
    \centering
  	\def\svgwidth{0.8\linewidth}
	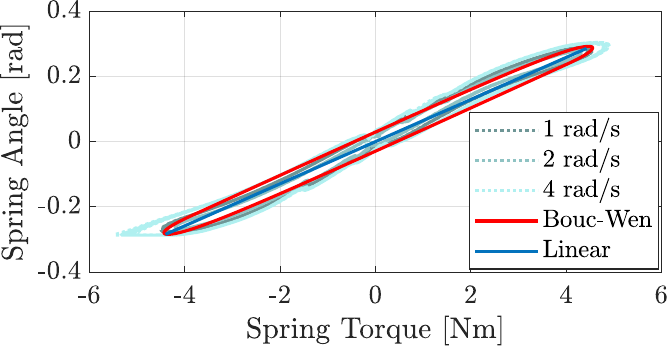 
  \vspace{-0.3cm}
    \caption{Linearized stiffness model combined with hysteresis loss.} %
 \label{fig:stiffness_model}
 \vspace{-0.7cm}
\end{figure}
\textbf{Embedded Controller Firmware.} Within the $\mu$C, a programmable multilevel cascaded controller runs and controls the actuators as shown in Fig.~\ref{fig:prototype_section_view}(e). This controller provides ease of use for presetting modes (e.g. initial position adjustment, calibration, etc.) or running modes such as those required by our optimal control trajectory generator. The inner loop (yellow) runs a current controller using PI-based Field Oriented Control \cite{pozof2022bldc}, the middle loop (cyan) is a PI-based speed controller with anti-windup, and the outer loop (green) is P-based position control. Three discrete estimators are used to estimate $\dot \theta_r$, $\dot q$, and $\dot \psi$. A low pass filter (LPF) is also implemented to filter the acquired $\tau_s$ signal. $\theta$ and $\dot \theta$ are calculated using the gearbox ratio $g_r$, of $\mathbb{M}^{M}$.  All control parameters, proportional and integral gains, cut-off frequency, etc. are sent to the system from the main controller via the EtherCAT bus. Both the outer and middle loops can be bypassed with $\theta^{ctrl}_{en}$ and $\dot \theta^{ctrl}_{en}$, respectively. the system can be controlled by an external reference signal via the EtherCAT bus when required. All measured and estimated variables such as $\theta$, $\dot \theta$, $q$, $\dot q$, $\psi$, $\dot \psi$, $i_q$, $\tau$, etc. are provided to the main controller via the EtherCAT bus.

\subsection{Operating Modes}
\label{sec:operating_modes}
As given in Tab.~\ref{tab:modes}, up to four operating modes can be used with the given configuration of our testbed.  
\begin{itemize}[leftmargin=*]
    \item \textbf{DEC} Mode ($p=0$) has both clutches disengaged, which means that no energy can be transferred to or from $\mathbb{L}^{M}$. Additionally, if energy was previously stored in $\mathbb{S}^{M}$, it gets dissipated by a free-unwinding spring.
    \item \textbf{SEA} Mode ($p=1$) has only $\mathbb{L}^{M}$ engaged, which means that energy can now be transferred from $\mathbb{M}^{M}$ to $\mathbb{L}^{M}$ via $\mathbb{S}^{M}$, creating an effective SEA system.
    \item \textbf{STG} Mode ($p=2$) has only the break engaged, which means that now energy can be stored in $\mathbb{S}^{M}$ without being interfered with movement $\mathbb{L}^{M}$.
    \item \textbf{BRK} Mode ($p=3$) has both modules engaged; thus, energy can be stored in $\mathbb{S}^{M}$ while $\mathbb{L}^{M}$ is forced to stop or hold in position. 
\end{itemize}

These four operation modes, compared to the only two ideal modes ($p=1,2$) presented in~\cite{ossadnik2022BSA} (due to the reinterpretation of the switch-and-hold mechanism), allow us not only to operate the system as a BSA but also gives us additional versatility while operating. %

\input{tables/parameters}
\begin{figure*}[ht]
    \vspace{0.2cm}
    \centering
  	\def\svgwidth{0.8\linewidth}
	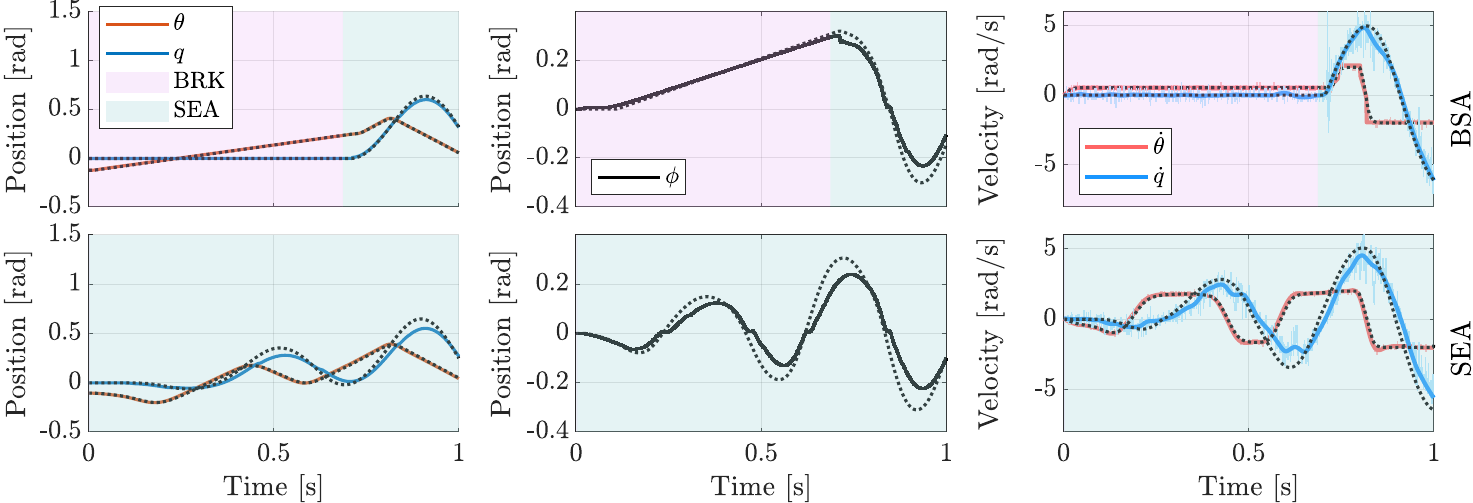
 \vspace{-0.3cm}
    \caption{Experimental results - \textit{Maximization of the link velocity at a final time $t_f = 1$s}. The top and bottom rows show the BSA and SEA cases, respectively. The background colour indicates the operation mode (lilac for BRK and turquoise for SEA). All plots show the acquired data (solid line) and simulation data (dotted line). The left plots show the changes in motor angle $\theta$ (red) and link angle $q$ (blue). The middle plots show the change in spring angle $\phi$ (black). the right plots show motor speeds $\dot \theta$ (red) and link speed $\dot q$ data: raw (light blue) and filtered (blue). The data were filtered by robust quadratic regression using Matlab's smoothdata function.}
    \label{fig:exp1}
    \vspace{-0.6cm}
\end{figure*}
\subsection{Identification}
\label{sec:identification}
As mentioned in Sec. \ref{sec:modelling} BSA system includes several independent systems with performance-effecting parameters, although most of these parameters can be obtained directly from manufacturers and CAD models, some need further experimentation to be identified. In our current system, the delay between the control comment and the physical engagement of the clutch and the exact hysteresis parameters of the $\mathbb{S}^{M}$ are needed to be identified.

\textbf{Clutch Delay.} 
\new{Engaging and disengaging tests were done to find the electromechanical delays for $\mathbb{B}^{M}$ and $\mathbb{C}^{M}$. For $\mathbb{C}^{M}$, we start in DEC and set $\mathbb{L}^{M}$ manually to $q = \frac{\pi}{2}$ rad. BRK was engaged to hold the system in place before moving to STG.}
$\mathbb{L}^{M}$ drops due to gravity and the mode is changed back to BRK when $q \le 0$. For $\mathbb{B}^{M}$, \new{we start in DEC and} set $\mathbb{M}^{M}$ in speed control with a relatively low set-point $\dot \omega^d = 0.5$ rad/s starting at $\theta = 0$. We switch to STG mode between $\theta = [0.05,0.25]$ rad, considering the maximum spring deflection, $\phi_{\mathrm{max}} =  0.3$ and then back to DEC when $\dot \omega^d = 0$ rad/s.
\new{Offline analysis evaluated $c_1$ and $c_2$ against $\tau$ and $\dot q$, showing $\mathbb{B}^{M}$ and $\mathbb{C}^{M}$ delays of 22 and 23 ms, respectively.}

\textbf{Stiffness.} %
The stiffness model for the integrated spring mechanism is provided in~\cite{compact_sea2009}. However, the effects of the loading rate and the loading cycle have not been investigated. Due to the system's design, it is susceptible to spring buckling and bending, which can cause a change in stiffness at different loading rates~\cite{mcy_sea_2021}. In order to identify the characteristics of our $\mathbb{S}^{M}$, a series of experiments were conducted. The $\mathbb{M}^{M}$ is set in speed control while $\mathbb{B}^{M}$ is engaged. The $\dot \theta^d$ is alternated between $\pm \dot{\theta}^{set}$ rad/s when it reaches the limits $\theta^{lim} = [-0.29,0.29]$ rad for 10 consecutive cycles in order to load and unload the $\mathbb{S}^{M}$ while $\dot{\theta}^{set} = {1,2,3,4}$ rad/s  in each experiment, respectively. We derived both a linear stiffness model and a model including hysteresis losses. The derived spring characteristic can be seen in Fig. \ref{fig:stiffness_model}.

%% file: figures/Implementation_Model.pdf_tex
\begingroup%
  \makeatletter%
  \providecommand\color[2][]{%
    \errmessage{(Inkscape) Color is used for the text in Inkscape, but the package 'color.sty' is not loaded}%
    \renewcommand\color[2][]{}%
  }%
  \providecommand\transparent[1]{%
    \errmessage{(Inkscape) Transparency is used (non-zero) for the text in Inkscape, but the package 'transparent.sty' is not loaded}%
    \renewcommand\transparent[1]{}%
  }%
  \providecommand\rotatebox[2]{#2}%
  \newcommand*\fsize{\dimexpr\f@size pt\relax}%
  \newcommand*\lineheight[1]{\fontsize{\fsize}{#1\fsize}\selectfont}%
  \ifx\svgwidth\undefined%
    \setlength{\unitlength}{378.98766225bp}%
    \ifx\svgscale\undefined%
      \relax%
    \else%
      \setlength{\unitlength}{\unitlength * \real{\svgscale}}%
    \fi%
  \else%
    \setlength{\unitlength}{\svgwidth}%
  \fi%
  \global\let\svgwidth\undefined%
  \global\let\svgscale\undefined%
  \makeatother%
  \begin{picture}(1,0.17478805)%
    \lineheight{1}%
    \setlength\tabcolsep{0pt}%
    \put(0,0){\includegraphics[width=\unitlength,page=1]{./figures/Implementation_Model.pdf}}%
    \put(0.01708237,0.09301664){\color[rgb]{0,0,0}\makebox(0,0)[lt]{\lineheight{1.25}\smash{\begin{tabular}[t]{l}$M$\end{tabular}}}}%
    \put(0.12377284,0.09406314){\color[rgb]{0,0,0}\makebox(0,0)[lt]{\lineheight{1.25}\smash{\begin{tabular}[t]{l}$S_i$\end{tabular}}}}%
    \put(0,0){\includegraphics[width=\unitlength,page=2]{./figures/Implementation_Model.pdf}}%
    \put(0.30229677,0.09572708){\color[rgb]{0,0,0}\makebox(0,0)[lt]{\lineheight{1.25}\smash{\begin{tabular}[t]{l}$S_o$\end{tabular}}}}%
    \put(0.43041593,0.13899008){\color[rgb]{0,0,0}\makebox(0,0)[lt]{\lineheight{1.25}\smash{\begin{tabular}[t]{l}$T_i$\end{tabular}}}}%
    \put(0.53207581,0.13940607){\color[rgb]{0,0,0}\makebox(0,0)[lt]{\lineheight{1.25}\smash{\begin{tabular}[t]{l}$T_o$\end{tabular}}}}%
    \put(0.59521702,0.09419325){\color[rgb]{0,0,0}\makebox(0,0)[lt]{\lineheight{1.25}\smash{\begin{tabular}[t]{l}$B$\end{tabular}}}}%
    \put(0.7518315,0.09344474){\color[rgb]{0,0,0}\makebox(0,0)[lt]{\lineheight{1.25}\smash{\begin{tabular}[t]{l}$C$\end{tabular}}}}%
    \put(0.93693221,0.09301664){\color[rgb]{0,0,0}\makebox(0,0)[lt]{\lineheight{1.25}\smash{\begin{tabular}[t]{l}$L$\end{tabular}}}}%
  \end{picture}%
\endgroup%

%% file: plots/hysteresis_plot.pdf_tex
\begingroup%
  \makeatletter%
  \providecommand\color[2][]{%
    \errmessage{(Inkscape) Color is used for the text in Inkscape, but the package 'color.sty' is not loaded}%
    \renewcommand\color[2][]{}%
  }%
  \providecommand\transparent[1]{%
    \errmessage{(Inkscape) Transparency is used (non-zero) for the text in Inkscape, but the package 'transparent.sty' is not loaded}%
    \renewcommand\transparent[1]{}%
  }%
  \providecommand\rotatebox[2]{#2}%
  \newcommand*\fsize{\dimexpr\f@size pt\relax}%
  \newcommand*\lineheight[1]{\fontsize{\fsize}{#1\fsize}\selectfont}%
  \ifx\svgwidth\undefined%
    \setlength{\unitlength}{320.02574158bp}%
    \ifx\svgscale\undefined%
      \relax%
    \else%
      \setlength{\unitlength}{\unitlength * \real{\svgscale}}%
    \fi%
  \else%
    \setlength{\unitlength}{\svgwidth}%
  \fi%
  \global\let\svgwidth\undefined%
  \global\let\svgscale\undefined%
  \makeatother%
  \begin{picture}(1,0.51805386)%
    \lineheight{1}%
    \setlength\tabcolsep{0pt}%
    \put(0,0){\includegraphics[width=\unitlength,page=1]{./plots/hysteresis_plot.pdf}}%
  \end{picture}%
\endgroup%

%% file: tables/parameters.tex
\begin{table}[t]
\centering
\vspace{0.2cm}
        \caption{System parameters of the prototype with corresponding high-level counterparts.} %
    \label{tab:parameters}
      \vspace{-0.3cm}
    \begin{tabular}{|l|cc|c|c|}
    \hline
\multicolumn{1}{|c|}{\begin{tabular}[c]{@{}c@{}}Low-Level\\ Parameters\end{tabular}}	& \multicolumn{1}{c|}{Sym.}	& \begin{tabular}[c]{@{}c@{}}High-Level\\ Param.\end{tabular}	& Unit		& Value		\\ \hline
Motor Inertia		    &\multicolumn{1}{c|}{$M$}     & \multirow{2}{*}{$J_m$}  & $\mathrm{kgm^2}$	& 3.7e-07     \\ \cline{1-2} \cline{4-5} 
Spring input inertia	&\multicolumn{1}{c|}{$S_i$}   &		                 & $\mathrm{kgm^2}$	& 5.3e-04     \\ \hline
Spring output inertia	&\multicolumn{1}{c|}{$S_o$}   & \multirow{5}{*}{$J_s$}	 & $\mathrm{kgm^2}$	& 5.8e-03     \\ \cline{1-2} \cline{4-5} 
T/S input inertia	    &\multicolumn{1}{c|}{$T_i$}   &						 & $\mathrm{kgm^2}$	& 6.4e-04      \\ \cline{1-2} \cline{4-5} 
T/S output inertia	    &\multicolumn{1}{c|}{$T_o$}   &                         & $\mathrm{kgm^2}$	& 1.4e-04\\ \cline{1-2} \cline{4-5} 
Brake Inertia           &\multicolumn{1}{c|}{$B$}      &                        & $\mathrm{kgm^2}$	& 7.0-04        \\ \cline{1-2} \cline{4-5} 
Clutch Inertia          &\multicolumn{1}{c|}{$C_i$}    &                           & $\mathrm{kgm^2}$ & 1.0e-03\\ \hline
Link Inertia            &\multicolumn{1}{c|}{$L$}      & {$J_l$}               &  $\mathrm{kgm^2}$	& 5.2e-01 \\ \hline 
Link Mass               &\multicolumn{2}{c|}{$L_m$}    &  $\mathrm{kg}$	   & 1.2 \\ \hline
Link CoM                &\multicolumn{2}{c|}{$L_{cm}$} &  $\mathrm{m}$	& 0.2 \\ \hline
Link  Coulomb friction  &\multicolumn{2}{c|}{$\tau_{C, q}$} &  $\mathrm{Nm}$	& 0.28 \\ \hline
Link viscous damping    &\multicolumn{2}{c|}{$d_{q}$} &  $\mathrm{kgm^2/s}$	& 0.08\\ \hline
Spring Coulomb fric.    &\multicolumn{2}{c|}{$\tau_{C, \psi}$} &  $\mathrm{Nm}$	& 0.11 \\ \hline
Spring viscous damp.    &\multicolumn{2}{c|}{$d_{\psi}$} &  $\mathrm{kgm^2/s}$	& 0.04 \\ \hline
Motor Torque            & \multicolumn{2}{c|}{$\tau_m$}    & $\mathrm{Nm}$	& $\pm 5$ \\ \hline
Joint Angle Range       & \multicolumn{2}{c|}{$\theta$} & $\mathrm{rad}$	& $\pm 1.2$         \\ \hline
Max Spring Defl.        & \multicolumn{2}{c|}{$\phi_{\mathrm{max}}$} & $\mathrm{rad}$	& $\pm 0.3$   \\ \hline
Max Spring Torque       & \multicolumn{2}{c|}{$\tau_{s, \mathrm{max}}$} & $\mathrm{Nm}$	& $\pm 4.5$   \\ \hline
Spring Stiffness        & \multicolumn{2}{c|}{$K$} &$\mathrm{Nm/rad}$& $15$    \\ \hline
\multirow{3}{*}{Spring Hysteresis}  & \multicolumn{2}{c|}{$\mathrm{\alpha}$}& \multirow{3}{*}{-} & $0.08$ \\ \cline{2-3}\cline{5-5}
                                    & \multicolumn{2}{c|}{$\mathrm{\beta}$} &   & \multicolumn{1}{c|}{$2.00$}  \\ \cline{2-3}\cline{5-5}
                                    & \multicolumn{2}{c|}{$\mathrm{\gamma}$} &   & \multicolumn{1}{c|}{$0.60$}  \\ \hline
Current Control $K_p$   & \multicolumn{2}{c|}{$K_p^i$} & $\mathrm{V/A}$ & 1-0e-01 \\ \hline
Current Control $T_i$   & \multicolumn{2}{c|}{$T_i^i$} & $\mathrm{Vs/A}$ &  3.6e03 \\ \hline
Speed Control $K_p$     & \multicolumn{2}{c|}{$K_p^{\dot \theta}$} & $\mathrm{rad/As}$ & 4.0e0 \\ \hline
Speed Control $T_i$     & \multicolumn{2}{c|}{$T_i^{\dot \theta}$} & $\mathrm{rad/A}$ &  1.0e02 \\ \hline
Position Control $K_p$  & \multicolumn{2}{c|}{$K_p^\theta$} & $\mathrm{1/s}$ &  6.0e01 \\ \hline
Motor Gear ratio        & \multicolumn{2}{c|}{$g_r$} & - &  $1:108$ \\ \hline
T/S LPF frequency       & \multicolumn{2}{c|}{$f^\tau_c$} & $\mathrm{Hz}$ &  1.0e03 \\ \hline

    \end{tabular}

    \vspace{-0.6cm}
\end{table}

%% file: plots/exp_1sec.pdf_tex
\begingroup%
  \makeatletter%
  \providecommand\color[2][]{%
    \errmessage{(Inkscape) Color is used for the text in Inkscape, but the package 'color.sty' is not loaded}%
    \renewcommand\color[2][]{}%
  }%
  \providecommand\transparent[1]{%
    \errmessage{(Inkscape) Transparency is used (non-zero) for the text in Inkscape, but the package 'transparent.sty' is not loaded}%
    \renewcommand\transparent[1]{}%
  }%
  \providecommand\rotatebox[2]{#2}%
  \newcommand*\fsize{\dimexpr\f@size pt\relax}%
  \newcommand*\lineheight[1]{\fontsize{\fsize}{#1\fsize}\selectfont}%
  \ifx\svgwidth\undefined%
    \setlength{\unitlength}{706.94586182bp}%
    \ifx\svgscale\undefined%
      \relax%
    \else%
      \setlength{\unitlength}{\unitlength * \real{\svgscale}}%
    \fi%
  \else%
    \setlength{\unitlength}{\svgwidth}%
  \fi%
  \global\let\svgwidth\undefined%
  \global\let\svgscale\undefined%
  \makeatother%
  \begin{picture}(1,0.34099352)%
    \lineheight{1}%
    \setlength\tabcolsep{0pt}%
    \put(0,0){\includegraphics[width=\unitlength,page=1]{./plots/exp_1sec.pdf}}%
  \end{picture}%
\endgroup%

%% file: 04_evaluation.tex
\section{Evaluation}
\label{sec:evaluation}
\new{To assess actuator performance, we conduct a series of experiments exploring various BSA modes. These experiments involve solving an optimal control problem (OCP), which we detail in the following section before describing and discussing the conducted experiments.}
\vspace{-0.1cm} 
\begin{figure}[t!]
    \centering
  	\def\svgwidth{0.9\linewidth}
	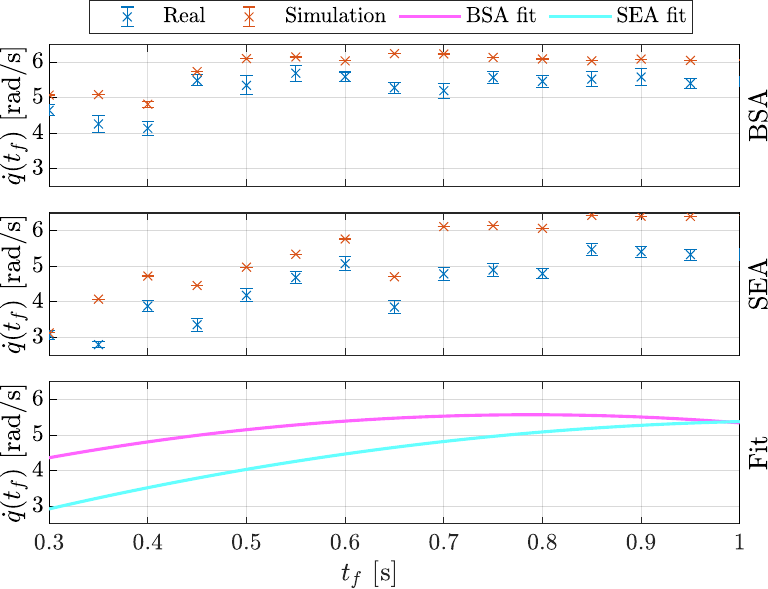 
 \vspace{-0.35cm}
    \caption{Statistical evaluation of the \textit{Maximization of $\dot q$ for varying final times $t_f$} experiment. The acquired data from the system (blue) and the online simulation data (red) are shown for the BSA (top) and SEA (middle) cases. A polynomial fit (bottom) of the mean velocities of the experimental data is shown. }
    \label{fig:exp1_stat}
    \vspace{0.1cm}
    \def\svgwidth{0.9\linewidth}
	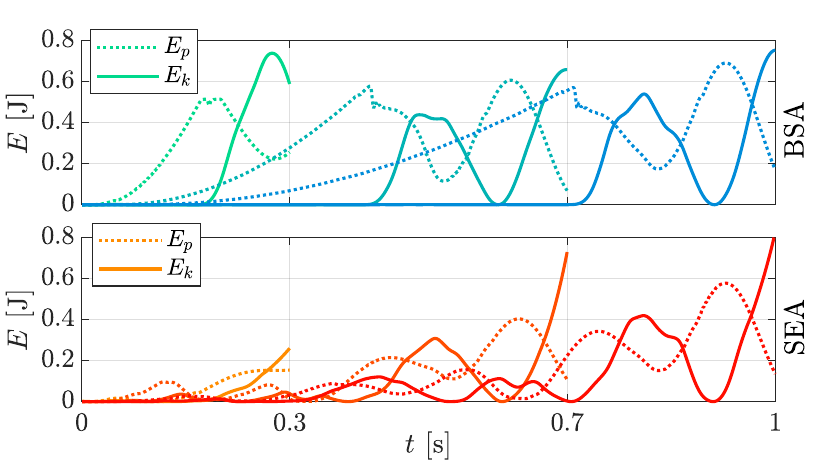 
 \vspace{-0.35cm}
    \caption{Evolution of the potential $E_p$ and kinetic $E_k$ energy for the \textit{Maximization of $\dot q$ for varying final times $t_f$} experiment. The plots are the superimposed cases of $t_f=\{0.3,0.7,1\}$s. } 
    \label{fig:energy}
     \vspace{-0.5cm} 
\end{figure}
\begin{figure*}[ht]
    \vspace{0.2cm}
    \centering
  	\def\svgwidth{0.8\linewidth}
	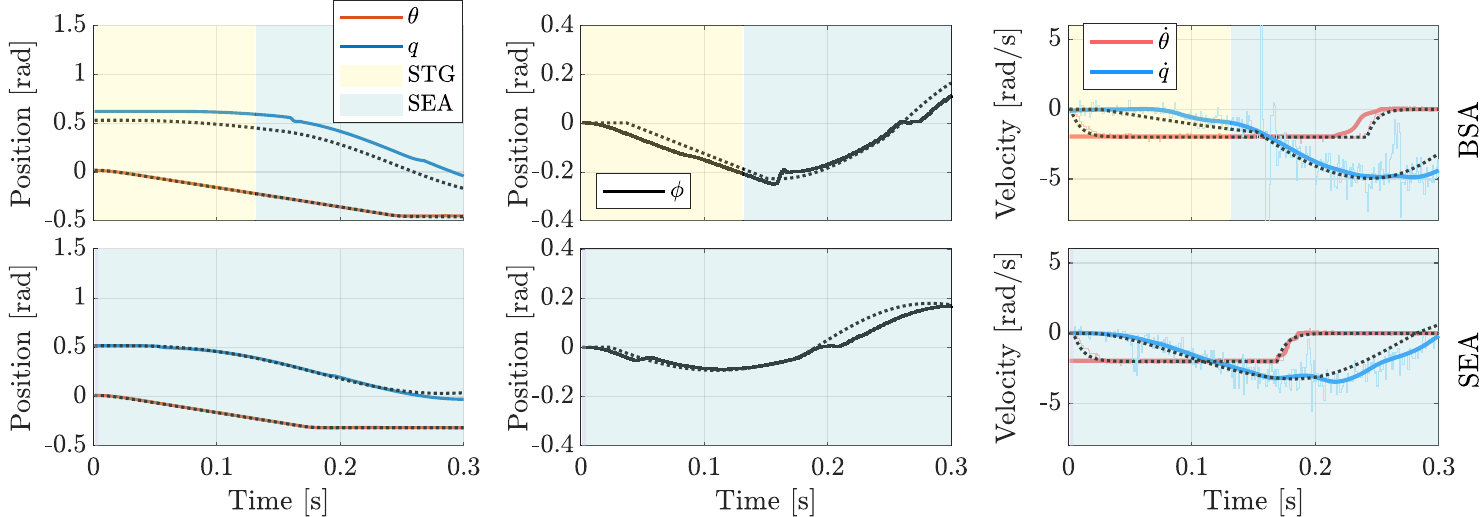 
    \caption{Experimental results - \textit{Maximization of $\dot q$ for an initial position $q_0 = 30$} deg. The top and bottom rows show the BSA and SEA cases, respectively. The background colour indicates the operation mode (yellow for STG and turquoise for SEA). All plots show the acquired data (solid line) and simulation data (dotted line). The left plots show the changes in motor angle $\theta$ (red) and link angle $q$ (blue). The middle plots show the change in spring angle $\phi$ (black). The right plots show motor speeds $\dot \theta$ (red) and link speed $\dot q$ data: raw (light blue) and filtered (blue).}
    \label{fig:exp2}
    \vspace{-0.6cm}
\end{figure*}
\subsection{Optimal Control} \label{sec:optimal_control} 
\new{The actuator dynamics' hybrid nature requires special treatment. We deal with it through a multi-phase optimization approach, where the mode sequence is defined a priori by the user. Using a direct collocation method \cite{bertsekas1997nonlinear}, we consider the time-series of states $\vect{x}(t)$, control input $u(t)$, and phase duration $T_p$ as decision variables. The reset map is evaluated at each $T_p$ to initiate optimization for the subsequent phase.}
We define the sets $\mathcal{X}$ and $\mathcal{U}$ as the admissible sets for state and control inputs. The OCP can now be formulated as
\begin{equation}
\begin{gathered}
\underset{\vect{x}(t), u(t), T_p}{\mathsf{min}}  \text{\;\;} \mathcal{J}(\vect{x}(t), u(t)) \\
\text{s.t. \quad} \dot{\vect{x}}(t) = \vect{f}_p(\vect{x}(t), u(t)),\;\; t \leq T_p  \\
\vect{x}_p^{+}(t) = \vect{g}_p(\vect{x}^{-}(t)),\;\; t = T_p  \\
\vect{x}(t) \in \mathcal{X}, \; u(t) \in \mathcal{U} ,
\end{gathered}
\end{equation}
\noindent where $\mathcal{J}(\vect{x}(t), u(t))$ is the cost function. As our main performance metric, we select the maximum reachable link velocity $\dot q$ at a defined final time $t_f$.
\begin{equation}
    \mathcal{J}(\vect{x}(t), u(t)) = -\dot q (t_f). \label{eq:cost_vel}
\end{equation}
The optimal control problem is formulated in CasADi \cite{Andersson2019} and solved using Ipopt \cite{wachter2006implementation}.
\subsection{Maximization of link velocity for varying final times $t_f$}
\new{In the first set of experiments, we assess the actuator's ability for explosive motions. Beginning from equilibrium at $\vect{\xi} = \vect{0}$, trajectories are optimized to maximize link velocity $\dot{q}$ within fixed final times $t_f \in [0; 1.0]$ s. The sequence \mbox{BRK $\rightarrow$ SEA} is encoded in the optimization.}
This way, the spring can be loaded with maximum speed and, when the brake is released, the stored potential energy can be quickly converted into kinetic energy. The BSA trajectories for $t_f = 1$s are shown in Fig. \ref{fig:exp1}. After loading the spring, at around $t = 0.7$s, the brake is released and the motor works in liaison with the spring to first accelerate in one direction, then in the other (essentially doing a single ``bang”).\\
On the other hand, if only the SEA mode is used, the actuator employs a resonant excitation strategy -- using a bang-bang-like signal which is typically associated with this type of actuator \cite{haddadin2009kick}. In this particular run, the BSA reaches a velocity of 6.11 rad/s whereas the SEA reaches 5.58 rad/s.
\new{Fig. \ref{fig:exp1_stat} depicts the statistical analysis of the experiment. With 10 repetitions for each $t_f$, the plot illustrates the mean and standard deviation for both the actual system (blue) and online simulation (red). Consistently, for $t_f > 0.45$ s, BSA achieves final velocities above $\dot{q} = 5$ rad/s. While maintaining good performance at higher $t_f$ values, SEA falls short in achieving comparable velocities at small $t_f$.\footnote{ Both BSA and SEA reach their peak velocity at some $t_f < 1$s. In this case, the optimal control algorithm found a slightly better local minimum.} Notably, for $t_f < 0.4$ s, the BSA solutions lack time for bidirectional acceleration (previous ``single bang"), leading to slightly lower final $\dot q$. The choice of stiffness value and end-time could be suboptimal for the SEA. For optimal performance, the resonance frequency of the system has to be aligned with the designated time window. While the SEA would excel in this specific combination, its performance would be less satisfactory for every other time window.}
\begin{figure}[t]
    \centering
  	\def\svgwidth{0.9\linewidth}
	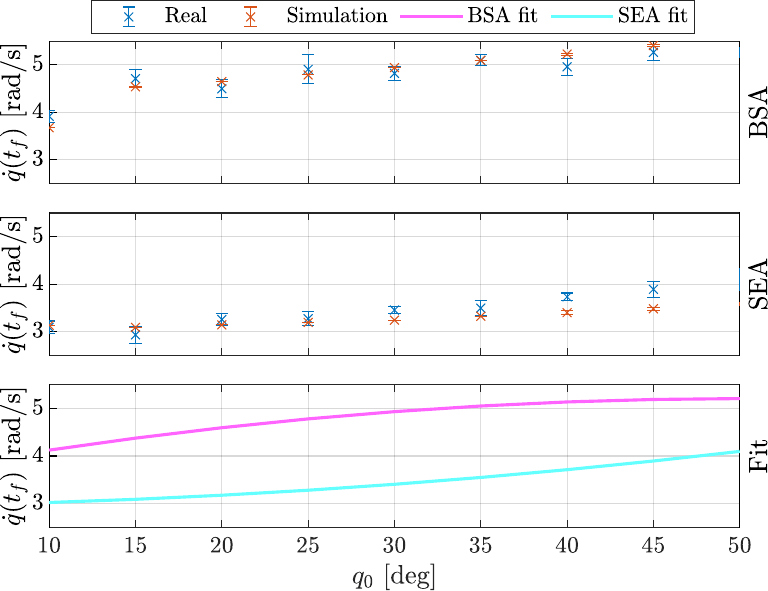
    \caption{Statistical evaluation of the \textit{Maximization of $\dot q$ for varying initial position $q_0$ } experiment. The acquired data from the system (blue) and the online simulation data (red) is shown for the BSA (top) and SEA (middle) cases. A polynomial fit (bottom) of the mean velocities of the experimental data is shown.} %
    \label{fig:exp2_stat}
\vspace{-0.5cm}
\end{figure}
\new{Next, we will examine the kinetic ($E_k$) and potential ($E_p$) energy evolution for BSA and SEA at $t_f={0.3,0.7,1}$\new{s}, as depicted in Fig.~\ref{fig:energy}.}
We can see how the BSA generates consistent launch sequences (slow build-up and rapid release). %
\new{When examining SEA signals, potential and kinetic energy oscillate for all end times, indicating resonant excitation.} The experiment confirms two of the main advantages of the BSA concept: 1) The timing of energy storage and release can be precisely controlled. 2) By avoiding resonant excitation, and thus lengthy swing-up motions, the actuator can reach a higher end-velocity in shorter periods of time.
\subsection{Maximization of the link velocity for varying initial positions $q_0$}
As described in \cite{Beckerle2017}, the initial position of the link can greatly influence the performance of an SEA. Therefore, we conduct another series of experiments, in which we again aim to maximize the link velocity, but start from an inclined angle $q_0 = 10 \dots 50$ deg. We generate optimal trajectories for a free end-time $t_f \in [0\;  0.5]$s. As before, each experiment is repeated 10 times.
An exemplary result for the BSA is shown in Fig. \ref{fig:exp2}. Here, the system is started from an initial angle of \mbox{$q_0 = 30$ deg}. First, while in STG mode, the actuator decouples the link and locks the spring. As the link is accelerated by gravity alone, the motor simultaneously loads the spring. At $t = 0.126$ s, the mode is changed to SEA. The potential energy stored in the spring is converted to kinetic energy, further accelerating the link.
In the SEA case, the motor moves with the link, but it cannot contribute much to accelerate it. This becomes also evident when viewing the spring deflection $\phi$, whose magnitude is relatively low compared to the BSA case. In the end, the BSA reaches a velocity of 4.88 rad/s, whereas the SEA only reaches 3.44 rad/s.
With the BSA's decoupling capability, we are not only able to precisely time the release of joint potential energy but also synchronize  its release with the gravity potential. In Fig. \ref{fig:exp2_stat}, the reached end velocity is plotted against the initial positions $q_0$. With increasing joint initial positions, the BSA also proportionally outperforms the SEA in terms of end velocity, exploiting the aforementioned synchronicity effect.

%% file: plots/exp1.pdf_tex
\begingroup%
  \makeatletter%
  \providecommand\color[2][]{%
    \errmessage{(Inkscape) Color is used for the text in Inkscape, but the package 'color.sty' is not loaded}%
    \renewcommand\color[2][]{}%
  }%
  \providecommand\transparent[1]{%
    \errmessage{(Inkscape) Transparency is used (non-zero) for the text in Inkscape, but the package 'transparent.sty' is not loaded}%
    \renewcommand\transparent[1]{}%
  }%
  \providecommand\rotatebox[2]{#2}%
  \newcommand*\fsize{\dimexpr\f@size pt\relax}%
  \newcommand*\lineheight[1]{\fontsize{\fsize}{#1\fsize}\selectfont}%
  \ifx\svgwidth\undefined%
    \setlength{\unitlength}{368.51662445bp}%
    \ifx\svgscale\undefined%
      \relax%
    \else%
      \setlength{\unitlength}{\unitlength * \real{\svgscale}}%
    \fi%
  \else%
    \setlength{\unitlength}{\svgwidth}%
  \fi%
  \global\let\svgwidth\undefined%
  \global\let\svgscale\undefined%
  \makeatother%
  \begin{picture}(1,0.7660345)%
    \lineheight{1}%
    \setlength\tabcolsep{0pt}%
    \put(0,0){\includegraphics[width=\unitlength,page=1]{./plots/exp1.pdf}}%
  \end{picture}%
\endgroup%

%% file: plots/plot_superimposed.pdf_tex
\begingroup%
  \makeatletter%
  \providecommand\color[2][]{%
    \errmessage{(Inkscape) Color is used for the text in Inkscape, but the package 'color.sty' is not loaded}%
    \renewcommand\color[2][]{}%
  }%
  \providecommand\transparent[1]{%
    \errmessage{(Inkscape) Transparency is used (non-zero) for the text in Inkscape, but the package 'transparent.sty' is not loaded}%
    \renewcommand\transparent[1]{}%
  }%
  \providecommand\rotatebox[2]{#2}%
  \newcommand*\fsize{\dimexpr\f@size pt\relax}%
  \newcommand*\lineheight[1]{\fontsize{\fsize}{#1\fsize}\selectfont}%
  \ifx\svgwidth\undefined%
    \setlength{\unitlength}{390.91360474bp}%
    \ifx\svgscale\undefined%
      \relax%
    \else%
      \setlength{\unitlength}{\unitlength * \real{\svgscale}}%
    \fi%
  \else%
    \setlength{\unitlength}{\svgwidth}%
  \fi%
  \global\let\svgwidth\undefined%
  \global\let\svgscale\undefined%
  \makeatother%
  \begin{picture}(1,0.56983698)%
    \lineheight{1}%
    \setlength\tabcolsep{0pt}%
    \put(0,0){\includegraphics[width=\unitlength,page=1]{./plots/plot_superimposed.pdf}}%
  \end{picture}%
\endgroup%

%% file: plots/exp30deg.pdf_tex
\begingroup%
  \makeatletter%
  \providecommand\color[2][]{%
    \errmessage{(Inkscape) Color is used for the text in Inkscape, but the package 'color.sty' is not loaded}%
    \renewcommand\color[2][]{}%
  }%
  \providecommand\transparent[1]{%
    \errmessage{(Inkscape) Transparency is used (non-zero) for the text in Inkscape, but the package 'transparent.sty' is not loaded}%
    \renewcommand\transparent[1]{}%
  }%
  \providecommand\rotatebox[2]{#2}%
  \newcommand*\fsize{\dimexpr\f@size pt\relax}%
  \newcommand*\lineheight[1]{\fontsize{\fsize}{#1\fsize}\selectfont}%
  \ifx\svgwidth\undefined%
    \setlength{\unitlength}{711.61752319bp}%
    \ifx\svgscale\undefined%
      \relax%
    \else%
      \setlength{\unitlength}{\unitlength * \real{\svgscale}}%
    \fi%
  \else%
    \setlength{\unitlength}{\svgwidth}%
  \fi%
  \global\let\svgwidth\undefined%
  \global\let\svgscale\undefined%
  \makeatother%
  \begin{picture}(1,0.34824578)%
    \lineheight{1}%
    \setlength\tabcolsep{0pt}%
    \put(0,0){\includegraphics[width=\unitlength,page=1]{./plots/exp30deg.pdf}}%
  \end{picture}%
\endgroup%

%% file: plots/exp2.pdf_tex
\begingroup%
  \makeatletter%
  \providecommand\color[2][]{%
    \errmessage{(Inkscape) Color is used for the text in Inkscape, but the package 'color.sty' is not loaded}%
    \renewcommand\color[2][]{}%
  }%
  \providecommand\transparent[1]{%
    \errmessage{(Inkscape) Transparency is used (non-zero) for the text in Inkscape, but the package 'transparent.sty' is not loaded}%
    \renewcommand\transparent[1]{}%
  }%
  \providecommand\rotatebox[2]{#2}%
  \newcommand*\fsize{\dimexpr\f@size pt\relax}%
  \newcommand*\lineheight[1]{\fontsize{\fsize}{#1\fsize}\selectfont}%
  \ifx\svgwidth\undefined%
    \setlength{\unitlength}{368.51662445bp}%
    \ifx\svgscale\undefined%
      \relax%
    \else%
      \setlength{\unitlength}{\unitlength * \real{\svgscale}}%
    \fi%
  \else%
    \setlength{\unitlength}{\svgwidth}%
  \fi%
  \global\let\svgwidth\undefined%
  \global\let\svgscale\undefined%
  \makeatother%
  \begin{picture}(1,0.7685809)%
    \lineheight{1}%
    \setlength\tabcolsep{0pt}%
    \put(0,0){\includegraphics[width=\unitlength,page=1]{./plots/exp2.pdf}}%
  \end{picture}%
\endgroup%

%% file: 05_conclusion.tex
\section{Conclusion}
\label{sec:conclusion}
\new{
In this study, we implemented and validated the novel BSA concept using a new re-configurable testbed. Confirming our earlier hypotheses, we demonstrated BSA's ability to: 1) achieve higher velocity than traditional SEAs in a limited time, and 2) generate clear launch sequences instead of harmonic oscillations. We also showcased BSA's capacity to synchronously release joint and gravity potential. \\
Future work involves designing and testing a multi-DoF BSA system. In this respect, we will especially focus on the effect of link decoupling which leads to underactuation and therefore potential limitations in terms of controllability. Furthermore, we aim to design closed-loop controllers allowing fully directed explosive motions. We believe that torque-controlled strategies are within reach now, too.}